\documentclass[3p, gulliver, 12pt, A4]{elsarticle}

\usepackage[utf8]{inputenc}
\usepackage{amsmath}
\usepackage[english]{babel}
\usepackage{mathtools}
\usepackage[version=4]{mhchem}
\usepackage{amsfonts}

\usepackage{appendix}

\usepackage{amssymb}
\usepackage{mathrsfs}
\usepackage{textcomp}
\usepackage{gensymb}

\usepackage{graphicx}
\usepackage[scriptsize]{caption}
\usepackage{caption}
\usepackage{subcaption}

\usepackage{wrapfig}

\usepackage{listings}

\usepackage{tikz}
\usepackage{tikz-3dplot}
\usetikzlibrary{arrows}
\usepackage{xparse}
\usetikzlibrary{3d,shapes,shadings}

\usepackage{enumitem}

\usepackage{footmisc}
\usepackage{color, soul}

\definecolor{lightorange}{rgb}{0.97,.69,0.3}

\definecolor{lightblue}{rgb}{.8,.95,1}

\usepackage{transparent}

\usepackage{eso-pic}

\usepackage{pdfpages}
\usepackage{float}

\usepackage{hyperref}
\hypersetup{
	colorlinks=true,
	citecolor=Purple,
	urlcolor=MyPink,      
	linkcolor= Blue,
	menucolor = Green,
}

\usepackage[makeroom]{cancel}
 \usepackage{wasysym}
 
\usepackage[font=footnotesize,labelfont=bf]{caption}

\begin{document}

\begin{frontmatter}
	
	\title{A hybrid approach for solving the gravitational $N$-body problem with Artificial Neural Networks}

	\author[1]{Veronica Saz Ulibarrena
		\corref{cor1}\fnref{fn1}}
	\cortext[cor1]{Corresponding author}
	\ead{ulibarrena@strw.leidenuniv.nl
	}
	\author[2]{Philipp~Horn}
	\ead{p.horn@tue.nl}
	\author[1]{Simon~Portegies~Zwart}
	\ead{spz@strw.leidenuniv.nl}
	
	\author[1]{\\Elena~Sellentin}
	\ead{sellentin@strw.leidenuniv.nl}
	\author[2]{Barry~Koren}
	\ead{b.koren@tue.nl}
	\author[1]{Maxwell~X.~Cai}
	\ead{cai@strw.leidenuniv.nl}

	\address[1]{Leiden Observatory, Leiden University, Niels Bohrweg 2, 2333 CA Leiden, The Netherlands}
	\address[2]{Eindhoven University of Technology, 5612 AZ Eindhoven, The Netherlands}

	\begin{abstract}
		Simulating the evolution of the gravitational $N$-body problem becomes extremely computationally expensive as $N$ increases since the problem complexity scales quadratically with the number of bodies. 
		In order to alleviate this problem, we study the use of Artificial Neural Networks (ANNs) to replace expensive parts of the integration of planetary systems.
		Neural networks that include physical knowledge have rapidly grown in popularity in the last few years, although few attempts have been made to 
		use them to 
		speed up the simulation of the motion of celestial bodies. For this purpose, we study the advantages and limitations of using Hamiltonian Neural Networks to replace computationally expensive parts of the numerical simulation of planetary systems, focusing on realistic configurations found in astrophysics. We compare the results of the numerical integration of a planetary system with asteroids with those obtained by a Hamiltonian Neural Network and a conventional Deep Neural Network, with special attention to understanding the challenges of this specific problem. Due to the non-linear nature of the gravitational equations of motion, errors in the integration propagate, which may lead to divergence from the reference solution. To increase the robustness of a method that uses neural networks, we propose a hybrid integrator that evaluates the prediction of the network and replaces it with the numerical solution if considered inaccurate.
		
		Hamiltonian Neural Networks can make predictions that resemble the behavior of symplectic integrators but are challenging to train and in our case fail when the inputs differ $\sim$7 orders of magnitude. In contrast, Deep Neural Networks are easy to train but fail to conserve energy, leading to fast divergence from the reference solution. The hybrid integrator designed to include the neural networks increases the reliability of the method and prevents large energy errors without increasing the computing cost significantly. For the problem at hand, the use of neural networks results in faster simulations when the number of asteroids is $\gtrsim$70.
		
	\end{abstract}
	
	
\end{frontmatter}

\section{Introduction}\label{Introduction} 
Planetary systems are a special case of the gravitational $N$-body problem in which a massive central star is orbited by multiple minor bodies, which include planets and asteroids among others. To model the evolution of planetary systems, it is necessary to know the gravitational interaction between the different bodies, which can be calculated using the equations derived by Newton \cite{NewtonPrincipia}. Unlike the calculation of the gravitational force, the equations of motion can only be solved analytically for two bodies using the relations derived by Kepler in 1609 \cite{kepler2015astronomia}. This means that when the system consists of three or more bodies, the equations need to be solved numerically with what we call the integrator. Hermite \cite{makino1991optimal} and Verlet \cite{verlet1967computer} integrators are frequently used for solving the general $N$-body problem, whereas others such as the Wisdom-Holman integrator \cite{wisdom1991symplectic} have been developed for the specific case of planetary systems.

Currently, the study of the evolution of $N$-body systems is limited by the large computational resources required to obtain an accurate\footnote{Accurate refers to solutions with low energy error.} solution \cite{greengard1990numerical, almojel2000implementation}. Newton’s equation of gravitation implies that the computational complexity of the problem scales with $N^2$. As a consequence, for multiple applications in astrophysics such as the evolution of globular clusters or asteroids around a star \cite{asteroids}, the large number of bodies in the system is one of the main reasons for the high computational cost.

Machine Learning (ML) is a tool with the potential to ameliorate this problem \cite{chevallier1998neural}. Although the applications of ML, or more precisely Artificial Neural Networks (ANNs), are scarce for the gravitational $N$-body problem \cite{tamayo, trani}, ANNs have recently demonstrated their potential in other fields \cite{doupe2019machine, basuchoudhary2017machine, mansfield2020predicting}. We study the efficiency of neural networks to replace computationally expensive parts of the integration of $N$-body systems for astrophysics applications.

Some studies have been carried out to apply ANNs to the two- and three-body gravitational problems to predict the future state of the system. For example, Breen et al.\ \cite{breen2020newton} in 2020 designed a Deep Neural Network (DNN) to replace the integration of the chaotic three-body problem. Their setup consists of three coplanar bodies of equal mass, with a zero initial velocity, which state is propagated in time using the arbitrary precise Brutus integrator developed by Boekholt and Portegies Zwart, 2015 \cite{Brutus}. The ANN receives as inputs the state of the particles at initial time $t_0$ and the simulation time $t$. In this simplified approach, the network is able to capture the complex motions of the three bodies, at a fraction of the computational expense.

Since the introduction of Physics-Informed Neural Networks (PINNs) in 2019 \cite{Raissi2019}, the popularity of neural networks with physics knowledge included has grown rapidly \cite{Lu2021,jin2020sympnets}. The claim is that the introduction of physical properties into the neural network allows for better predictions, better extrapolation capabilities, and less training data. So far, PINNs have not been applied to astrophysics problems. Following the idea of introducing physical knowledge into the neural network, Greydanus et al. \cite{greydanus2019hamiltonian} developed in 2019 Hamiltonian Neural Networks (HNNs) to 
address Hamiltonian mechanics within the network's architecture. To study the performance of their network, they use the gravitational two- and three-body problems as test cases. For the two-body problem, Greydanus et al. found that the HNN can predict the trajectories of the particles better than a baseline DNN. However, for the three-body problem, both networks fail to predict the trajectories. An alternative for HNNs was developed by Chen and Tao in 2021 \cite{chenGFNN}, denominated as Generating Function Neural Networks (GFNNs). They tested this approach on the two-body problem with similar inputs as in Greydanus et al. \cite{greydanus2019hamiltonian}. The comparison with other types of neural networks such as HNNs and SympNets \cite{jin2020sympnets} shows that GFNNs outperform the other methods for this particular test case. Although the results of Greydanus et al. \cite{greydanus2019hamiltonian}, and Chen and Tao \cite{chenGFNN} are promising, both references take the two- and three-body problems as test cases to demonstrate the performance of their neural networks. It is not yet certain that the introduction of physics into the neural network represents an advantage for more complicated problem configurations. For that reason, we study the advantages and disadvantages of HNNs when applied to {more realistic} astrophysics problems, in particular, the orbital evolution of celestial bodies.

We study the use of neural networks for the integration of planetary systems formed by two planets and up to 2,000 asteroids. Due to the popularity of physics-aware neural networks, we compare the results of direct numerical integration with the predictions of a network that includes physical knowledge (HNN) and a conventional Deep Neural Network (DNN). In \autoref{sub:hybrid}, we discuss the setup of a hybrid integrator that uses the neural network but switches to direct numerical integration when the former fails to produce sufficiently reliable answers. This method is faster than the classical integration and more accurate than the neural network. In \autoref{sec:ANNs}, we discuss the hyperparameter selection and the training results of the neural networks. In \autoref{subsec:validation}, we find the improvement in performance by the neural networks in the form of computation time as a function of the number of asteroids in the system, and in \autoref{subsec:results} we show the results of integrating a planetary system. 
The code is publicly available at \url{https://github.com/veronicasaz/PlanetarySystem_HNN}.

\section{Methodology}\label{sec:methodology}

\subsection{Numerical integration}

We consider a system of $N$ point masses interacting only via their Newtonian gravitational force. The gravitational force exerted on a body $i$, can be written as a function of mass ($m$), position vector ($\vec q$), and the universal gravitational constant ($G$) as

\begin{equation}\label{eq:Newton}
m_i \;\dfrac{d^2\vec q_i}{dt^2} = \sum_{j = 0,\;j \neq i}^{N-1} G \; \dfrac{m_i m_j}{\vert \vert \vec q_{ij}\vert \vert^3} \vec{q}_{ij}, \qquad \vec q_{ij} = \vec q_j - \vec q_i,
\end{equation}
where the indices $i$ and $j$ denote the celestial bodies.\\

Knowing the acceleration vector, the state of the system can be evolved in time using an integrator. Wisdom and Holman in 1991 \cite{wisdom1991symplectic} proposed a symplectic integrator for systems in which one body is much more massive than the others. In our case, we assume that the smaller bodies orbit this massive one and the barycenter of the system is located approximately at the center of the massive body. The other bodies orbit the barycenter in almost Keplerian trajectories. 

The Hamiltonian of the system is given by
\begin{equation}\label{eq:Hamiltonian}
\mathcal{H} = 
\sum_{i=0}^{N-1} \dfrac{\vert \vert \vec p_i \vert \vert^2}{2m_i} - G \sum_{i=0}^{N-2} m_i\sum_{j=i+1}^{N-1} \;\dfrac{m_j}{\vert \vert \vec q_j - \vec q_i \vert \vert},
\end{equation}
where $\vec p$ represents the linear momentum vector.

For planetary systems, Equation (\ref{eq:Hamiltonian}) can be split into two parts.  Due to the assumption of the Sun being at the barycenter, i = 0 is excluded from the following equations. The first one, the Keplerian part,
\begin{equation}\label{eq:H_kepler}
\mathcal{H}_{\text{Kepler}} = \sum_{i=1}^{N-1} \dfrac{\vert \vert \vec p_i \vert \vert^2}{2m_i} - G  m_0\sum_{j=1}^{N-1} \;\dfrac{m_j}{\vert \vert \vec q_j\vert \vert},
\end{equation}

contains the terms related to the kinetic energy of the bodies and the potential energy due to the central body (body 0). The second part called interactive part,

\begin{equation}\label{eq:H_inter}
\mathcal{H_{\text{inter}}} = -G \sum_{i=1}^{N-2} m_i\sum_{j=i+1}^{N-1} \;\dfrac{m_j}{\vert \vert \vec q_j - \vec q_i \vert \vert},
\end{equation}

contains the terms with the potential energy due to the mutual interaction between the orbiting bodies.

The Wisdom-Holman (WH) integrator first propagates the trajectory of the orbiting bodies without taking their mutual interaction into account by performing a Keplerian propagation around the central body. After that, the perturbing acceleration is calculated and converted to a correction of the velocity. 

Although Equations (\ref{eq:H_kepler}) and (\ref{eq:H_inter}) are expressed in heliocentric coordinates for clarity, WH's integrator uses Jacobian coordinates for parts of its integration, as explained in Wisdom and Holman \cite{wisdom1991symplectic}.

The computing time of the Keplerian propagation scales linearly with the number of bodies ($N$) as seen in Equation ({\ref{eq:H_kepler}}). In contrast, the interactive part (Equation (\ref{eq:H_inter})) scales quadratically with the number of bodies. Therefore, it is interesting to find methods to speed up the latter. We use ANNs to replace the interactive part to speed up the calculation of the mutual perturbations.

\subsection{Neural Network surrogates}\label{subsec:NNs}

In examples such as Breen et al. {\cite{breen2020newton}} and Greydanus et al. {\cite{greydanus2019hamiltonian}}, a neural network is used to replace the integrator. However, this approach falls short for many astrophysics applications. For example, for the case of a planetary system, the force exerted by the central body is orders of magnitude larger than the mutual forces exerted by the orbiting bodies. If a neural network is used to predict the future state of the system, it will fail to capture the smaller contributions of the planets. We propose a method in which the neural network is integrated into the numerical integration without losing information about the perturbations. We do so by calculating the Keplerian Hamiltonian $\mathcal{H}_{\text{Kepler}}$ analytically and the interactive Hamiltonian using a neural network $\mathcal{H}_{\text{inter}}$.

For systems in which energy is conserved, Hamiltonian Neural Networks (HNNs) constitute an attractive choice since the Hamiltonian of the system can be input as a physical constraint into its architecture. We therefore use HNNs to predict the interactive part of Equation (\ref{eq:Hamiltonian}) similarly to the Neural Interacting Hamiltonian (NIH) designed by Cai et al.\ \cite{cai2021neural}. We study the advantages and disadvantages of HNNs by comparing them to the numerical integration, which we consider the baseline, and to a conventional Deep Neural Network (DNN). 

\begin{figure}[h!]
	\centering
	\includegraphics[scale=1]{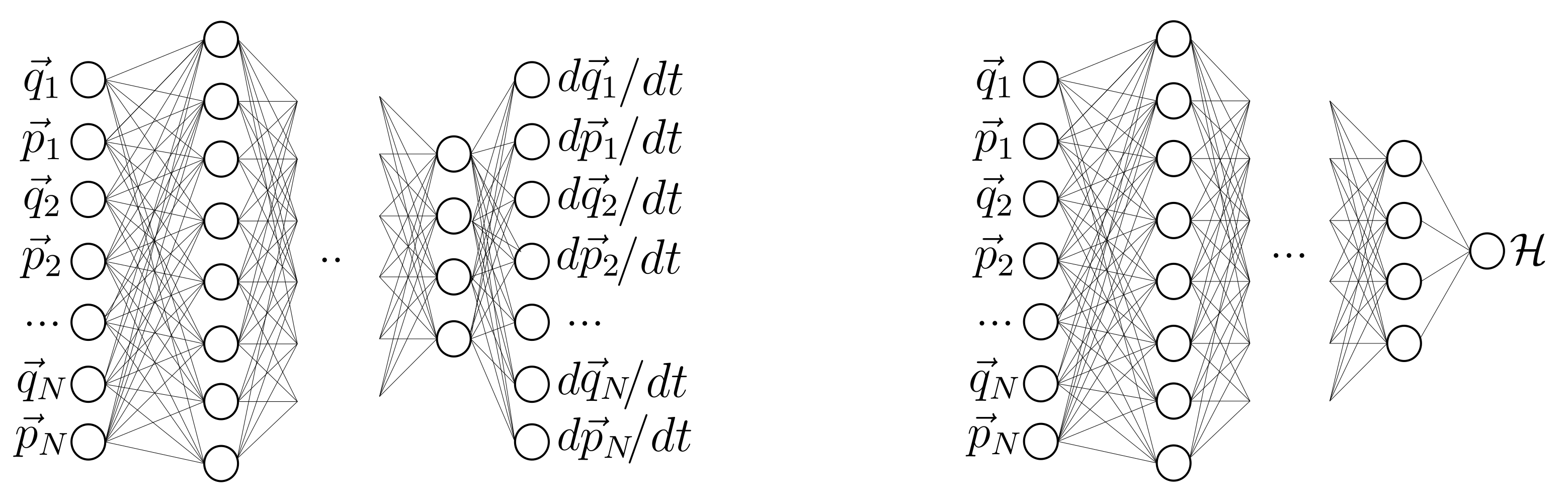}\\
	$ $\hspace{70pt}(a) \hspace{190pt} (b) \hspace{60pt}
	\caption[Schematic of a) a Deep Neural Network and b) a Hamiltonian Neural Network]{Schematic of a a) Deep Neural Network and a b) Hamiltonian Neural Network that predict the derivatives of the inputs with respect to time and the Hamiltonian of the system, respectively. The inputs for both are the position and linear momentum of all objects in the system.}
	\label{fig:HNNschematic}
\end{figure}

An HNN \cite{greydanus2019hamiltonian} receives as inputs the position and linear momentum of all the bodies in the system and outputs the Hamiltonian of the system. In {\autoref{fig:HNNschematic}} we show a comparison of an uninformed neural network (DNN) and a HNN.

With the output of the HNN and automatic differentiation, the derivatives of the inputs are calculated using Hamilton's equations:
\begin{equation}\label{eq:hamiltonianproperties}
-\dfrac{\partial \mathcal{H}}{\partial \vec q} = \dfrac{d\vec p}{dt}, \hspace{60pt}
\dfrac{\partial \mathcal{H}}{\partial \vec p} = \dfrac{d\vec q}{dt}.
\end{equation}
The derivatives are then used to compute the loss function during the training of the network.

Unlike in Greydanus et al. {\cite{greydanus2019hamiltonian}}, we use the neural networks for the calculation of the interactive Hamiltonian as expressed in Equation ({\ref{eq:H_inter}}). This Hamiltonian is only a function of the masses and positions of the different bodies, and the universal gravitational constant. This means that the neural networks from \autoref{fig:HNNschematic} can be simplified by eliminating the linear momentum from the inputs. Since the acceleration requires knowing the masses of the system, the inputs then become:
\begin{equation}
X = [m_1, \vec q_1, m_2, \vec q_2, ..., m_N, \vec q_N].
\end{equation}
Similarly, the outputs of the DNN are now reduced to the derivatives of the linear momentum with respect to time. By doing this, we achieve a substantial reduction in the number of parameters of the network. We will explain whether the symplectic structure of the integrator is conserved when using neural networks for the calculation of the interactive Hamiltonian in \autoref{subsec:symplectic}.

Taking into account this modification of the set of inputs and outputs, the loss function $\mathcal L$ for the HNN is the difference between the acceleration calculated using Newton's equation and the one obtained from differentiating the output of the HNN using Equation ({\ref{eq:hamiltonianproperties}}): 

\begin{equation}\label{eq:loss}
\mathcal{L_{\text{HNN}}(\theta)} = \dfrac{1}{M} \sum_{i=1}^M\left( -\dfrac{\partial \mathcal{H}_{\text{pred}}}{\partial \vec q} - \dfrac{d\vec p}{dt}\right)^2.
\end{equation}
In Equation ({\ref{eq:loss}}), $\theta$ represents the trainable parameters of the network, $\mathcal{H}_{pred}$ is the output of the HNN, and the gradients of $\mathcal H$ are obtained using automatic differentiation. $M$ is the number of samples for which the loss is being evaluated.

For the DNN, the inputs are the same as for the HNN and the derivatives of the inputs with respect to time are the outputs of the neural network. Therefore, the loss function is written as:

\begin{equation}\label{eq:lossDNN}
\mathcal{L}_{\text{DNN}}(\theta) = \dfrac{1}{M} \sum_{i=1}^M\left(
\left[\dfrac{d\vec p}{dt}\right]_{\text{pred}} - \dfrac{d\vec p}{dt}
\right)^2.
\end{equation}

\subsection{Hybrid numerical method}\label{sub:hybrid}

The use of neural networks to replace parts of the integration raises several challenges. Firstly, neural networks cannot be expected to be as accurate as the numerical calculation: the use of ANNs implies a loss in accuracy with the goal of improving computing speed. Secondly, since integration is a repetitive process in which the output of one time step is used as the input for the next one, errors propagate in time. In non-linear systems, this may quickly lead to unphysical solutions. In previous research trying to solve the $N$-body problem using neural networks, it is common to propagate over short time scales. This implies that the accumulation of errors is not relevant, but does not constitute a realistic case for astrophysics problems. To address this problem, we develop a hybrid method in which the prediction of the neural network is evaluated and replaced by the numerical solution if considered insufficiently accurate. 

Evaluating the accuracy of the prediction is not straight-forward since we want to avoid using Newton's equation. Therefore, we use as a measurement of accuracy the fact that accelerations should be fairly smooth in time. We evaluate the prediction of the network by comparing it to the prediction of the previous time step. Since the perturbations are expected to be rather smooth, we assume that a large difference between the acceleration predicted by the network at time $t_0+\Delta t$ and the acceleration at $t_{0}$ is an indication of either a poor prediction or a region with quick changes in the acceleration. In both cases, it is beneficial to calculate those steps numerically instead of relying on the neural network. Although accelerations are expected to vary smoothly in time, by using a numerical time integrator we need to account for the discretization error when setting the tolerance $R$ for this smoothness criterion. From now on, we use the term ``flag" when the prediction of the network is not accepted. We calculate the acceleration $\vec a^{\;(t)}$ at time $t = t_0 +\Delta t$ by numerical integration if

\begin{equation}\label{eq:hybridcriteria}
\dfrac{\vert \vert \vec a^{\;(t_0)} - \vec a^{\;(t)} \vert \vert}{\vert \vert\vec a^{\;(t_0)}\vert \vert + \varepsilon} > R.
\end{equation}
This criterion represents the relative difference between the previous acceleration and the current one. The addition of $\varepsilon = 1\times 10^{-11}$ prevents the denominator from becoming zero. We adopt $R = 0.3$ to achieve an accurate reproduction of the trajectory whereas higher values result in larger energy errors, as we show in \ref{appendix:hybrid}.  The value of $R$ should be chosen according to the specifications of the problem at hand. If computational speed is more important than accuracy, higher values of $R$ could be chosen, whereas if the focus is on accuracy, $R$ should be smaller. A value of 0.3 represents a strict case in which ensuring accuracy is considered more important than achieving a low computational cost.

\begin{figure}[h!]
	\centering
	\includegraphics[scale=0.7]{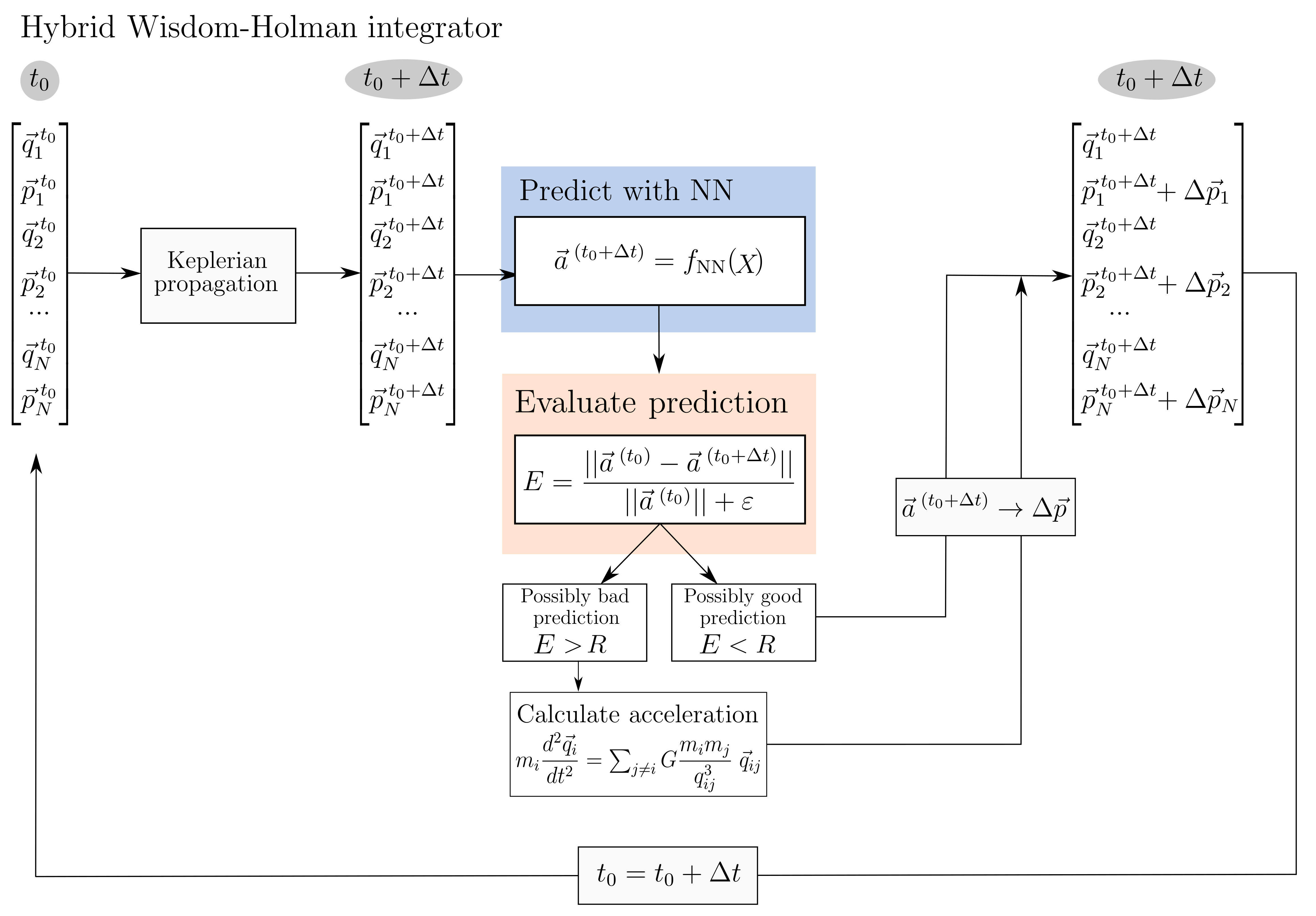}
	\caption[Schematic of the hybrid Wisdom-Holman integration]{Schematic of the hybrid Wisdom-Holman integration. The state of the system defined by $\vec p$ and $\vec q$ is propagated as a Keplerian trajectory. Then the Neural Network predicts the mutual perturbation between bodies. If the prediction is insufficiently smooth, the accelerations are calculated numerically using Equation ({\ref{eq:Newton}}). Finally, these accelerations are added as a change in linear momentum ($\Delta \vec p$). This process is repeated in the next time steps.}
	\label{fig:WHschematic}
\end{figure}

In \autoref{fig:WHschematic} we show the schematic diagram of the hybrid WH integrator. At time $t_0$, the state of each body is propagated a time step $\Delta t$, assuming that the particle is on a Keplerian trajectory. Afterward, the neural network ($f_{\text{NN}}$) calculates the perturbing accelerations for the given inputs ($X$). This prediction is evaluated and if considered insufficiently smooth according to the criterion defined in Equation (\ref{eq:hybridcriteria}), the accelerations are recalculated analytically. The perturbing accelerations are then converted into corrections in the velocity and the new state of the system is subsequently used as the starting point for the next time step. 

\subsection{Symplecticity of the integrator}\label{subsec:symplectic}

The original Wisdom-Holman integrator is a symplectic integrator {\cite{wisdom1991symplectic}}. Using a symplectic integrator is essential for long-term stability and energy conservation of Hamiltonian systems {\cite{Hairer2006}}. Some attempts have been made to conserve the symplectic structure of the integrator when using neural networks, such as in Zhu et al. \cite{zhu2020deep}. When using neural networks as surrogates for the calculation of the interactive part in the new hybrid integrator, it is beneficial to have the same symplectic structure as the original Wisdom-Holman integrator. 

The first part in the hybrid Wisdom-Holman integrator is the Keplerian propagation, which is the flow map of a Hamiltonian system and therefore a symplectic map. In the second part, the linear momentum vector is updated with accelerations either calculated by an ANN or using Newton's equation for the interactive part. In this step, the positions are always kept unchanged. If this update forms a symplectic map, the whole hybrid integrator becomes symplectic since concatenations of symplectic maps are again symplectic.

The update of the linear momentum vector only depends on the positions and not on the current momenta. So, the left-hand side of the symplectic condition

\begin{equation}
J_{\vec f}^T \begin{pmatrix} 0 & I \\ -I & 0 \end{pmatrix} J_{\vec f} = \begin{pmatrix} 0 & I \\ -I & 0 \end{pmatrix}
\end{equation}

of the Jacobian matrix $J_{\vec f}$ of a map $\vec f(\vec p, \vec q)$ simplifies to
\begin{align}
J_{\vec f}^T \begin{pmatrix} 0 & I \\ -I & 0 \end{pmatrix} J_{\vec f} & = \begin{pmatrix} I & 0 \\ \Delta t \left( \dfrac{\partial \vec a}{\partial \vec q} \right)^T & I \end{pmatrix} \begin{pmatrix} 0 & I \\ -I & 0 \end{pmatrix} \begin{pmatrix} I & \Delta t \left( \dfrac{\partial \vec a}{\partial \vec q} \right) \\ 0 & I \end{pmatrix} \\
& = \begin{pmatrix} I & 0 \\ \Delta t \left( \dfrac{\partial \vec a}{\partial \vec q} \right)^T & I \end{pmatrix} \begin{pmatrix} 0 & I \\ -I & -\Delta t \left( \dfrac{\partial \vec a}{\partial \vec q} \right) \end{pmatrix} \\
& = \begin{pmatrix} 0 & I \\ -I & \Delta t \left( \left( \dfrac{\partial \vec a}{\partial \vec q} \right)^T - \left( \dfrac{\partial \vec a}{\partial \vec q} \right) \right) \end{pmatrix} .
\end{align}
This implies that the Jacobian matrix of the calculated accelerations $\left( \frac{\partial \vec a}{\partial \vec q} \right)$ has to be symmetric.

If the accelerations are calculated using Newton's equation or using an HNN, they are the gradient of a scalar function, the Hamiltonian. This means that the Jacobian matrix of the accelerations is the Hessian matrix of the Hamiltonian, which is symmetric for continuous second derivatives. However, if a DNN is used in the hybrid integrator, no such statement can be made and the Jacobian matrix of the predicted accelerations can be non-symmetric.

Therefore, we can expect the energy-preserving characteristics of symplectic integrators to be present when using HNNs within the WH integrator but not when including DNNs. This result is investigated numerically in \autoref{subsec:results}.

\subsection{Problem setup}
\begin{figure}[h!] 
	\centering
	\begin{subfigure}{.45\textwidth}
		\centering
		\includegraphics[scale=0.2]{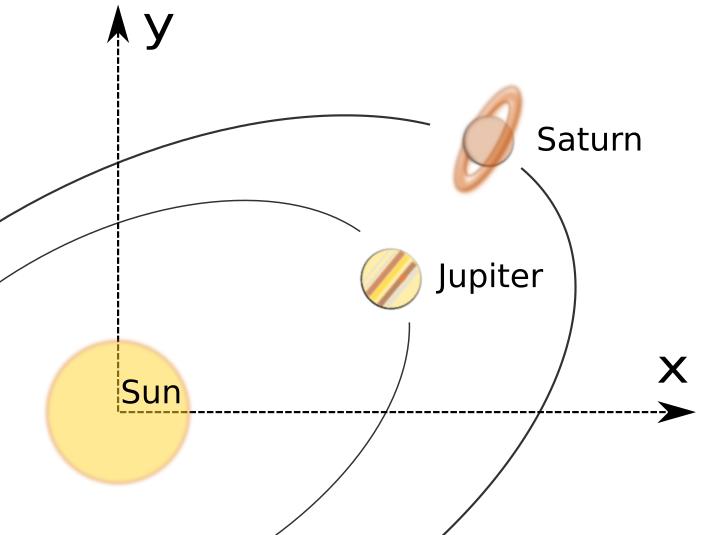}
		\caption{}
	\end{subfigure}
	\begin{subfigure}{.45\textwidth}
		\begin{minipage}{0.4\textwidth}
			\centering
			\includegraphics[scale=0.2]{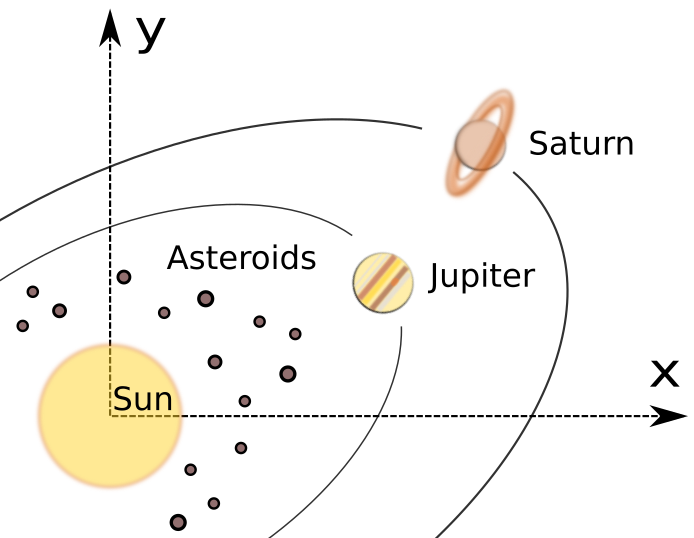}
			\caption{}
		\end{minipage}
	\end{subfigure}
	
	\caption[Schematic of the bodies in the two study cases]{Schematic of the problem setup. (a) First study case with the Sun, Jupiter, and Saturn. (b) Second study case with the Sun, Jupiter, Saturn, and asteroids located within Jupiter's orbit. }
	\label{fig:jupsat}
\end{figure}

We study two cases: the first one, Jupiter and Saturn orbiting the Sun, and the second one with a large number of asteroids added to the first case, as illustrated in  \autoref{fig:jupsat}.

For the first case in which only the Sun, Jupiter, and Saturn are studied (from now on referred to as SJS), the Hamiltonian of the system is given by Equation (\ref{eq:Hamiltonian}) for $N = 3$ with the Sun as $i=0$, Jupiter $i=1$, and Saturn $i=2$. The interactive part of the Hamiltonian corresponds to the interaction between Jupiter ($J$) and Saturn ($S$):
\begin{equation}
\mathcal{H_{\text{inter}}} = - G \dfrac{m_{J} m_{S}}
%{q_{J S}}
{\vert \vert \vec q_{J} - \vec q_{S}\vert \vert}.
\end{equation}

In this case, only one operation suffices to calculate the interactive part, and as a consequence, the use of ANNs will lead to a deceleration of the calculation. However, this setup constitutes an interesting study case. We set up the network for the inputs ($X$) to be the masses and positions of the two bodies and the output to be the Hamiltonian, as explained in \autoref{subsec:NNs}. Therefore, for the SJS case, the inputs are:
\begin{equation}
X_{\text{SJS}} = [m_J, \vec q_J, m_S, \vec q_S].
\end{equation}

In the second case (to which we refer as SJSa), we add $N_a$ asteroids in orbit around the massive central body. The Hamiltonian can again be calculated for the star, the two planets, and the asteroids with $N = 3 + N_a$. For example, when $N_a = 2$, the interactive Hamiltonian can be expressed as follows:
\begin{equation}\label{eq:HamiltonianJSa}
\mathcal{H}_{inter} = - G \dfrac{m_{J}m_{S}}{q_{{J}{S}}} - G \dfrac{m_1m_{J}}{q_{1{J}}} - G \dfrac{m_1m_{S}}{q_{1{S}}}- G \dfrac{m_2m_{J}}{q_{2{J}}} - G
\dfrac{m_2m_{S}}{q_{2{S}}} - G \dfrac{m_1m_2}{q_{12} },
\end{equation}
with $q$ representing the magnitude of $\vec q$.
Because asteroids are orders of magnitude less massive than the planets, it can be safely assumed that the mutual gravitational interaction between asteroids is negligible, and we therefore neglect the last term in Equation (\ref{eq:HamiltonianJSa}). We also assume that the effect of the asteroids on Jupiter and Saturn is negligible. In contrast, the effect of the planets on the asteroids cannot be neglected. To set up a neural network that predicts the perturbations on the asteroids due to the planets, we separate this interactive Hamiltonian for each of the asteroids. For asteroids 1 and 2, their interactive Hamiltonian is defined as:

\begin{align}
\mathcal{H}_1 & = - G \dfrac{m_{J} m_{S}}{q_{{J}{S}}} - G \dfrac{m_1m_{J}}{q_{1{J}}} - G \dfrac{m_1m_{S}}{q_{1{S}}},\\[1em]
\mathcal{H}_2 & = - G \dfrac{m_{J}m_{S}}{q_{{J}{S}}}- G \dfrac{m_2m_{J}}{q_{2{J}}} - G \dfrac{m_2m_{S}}{q_{2{S}}}.
\end{align}

We now set up the network such that the position and mass of each of the two asteroids correspond to one set of inputs. Therefore:
\begin{equation}
X_{\text{SJSa}} = [m_J, \vec q_J, m_S, \vec q_S, m_a, \vec q_a],
\end{equation}
where the subindex $a$ represents one of the $N_a$ asteroids. This choice of inputs allows the size of the neural network to be independent of the number of asteroids in the system, which implies that the same neural network can be used for any number of asteroids without retraining.

\section{Neural network results}\label{sec:ANNs}

In this section, we explain the creation of the training dataset, the choice of hyperparameters, and the training results for the Hamiltonian Neural Network and the Deep Neural Network. 

\subsection{Dataset}
\begin{figure}[h!]
	\centering
	\includegraphics[scale=0.4]{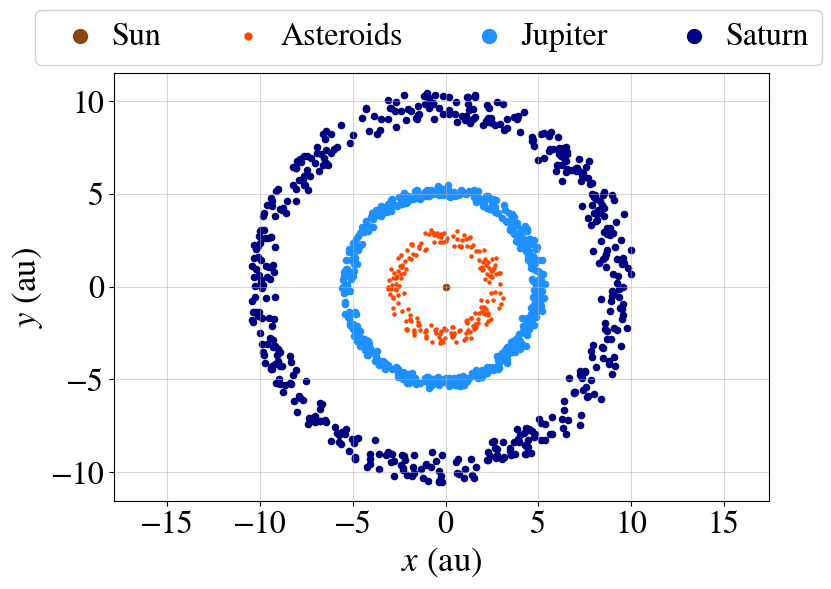}
	
	\caption[]{Distribution of $x$ and $y$ positions of the Sun, Jupiter, Saturn, and the asteroids in the training dataset.}
	\label{fig:datasettraj}
\end{figure}

We generate training and test datasets for each of the two cases: SJS and SJSa. The ranges of values can be found in  \autoref{table:dataset_inputs} of \ref{appendix:datasetparam}. From these, the initial conditions are chosen using Latin hypercube sampling \cite{latinhyper} and the simulations are run until the end time is reached. At each time step, the state of the system is saved as a training sample. Then, we verify if the dataset created covers the entire search space, i.e., if there are samples in the full range of true anomaly [$0^\circ- 360^\circ$), which is displayed in \autoref{fig:datasettraj}. 

With the time step and the end time in \ref{appendix:datasetparam}, the number of training samples is 3,000,000.  We randomly choose a fraction of these for the training. On an AMD Ryzen 9 5900hs, it takes  $\sim$80 min to generate this dataset. All experiments utilize this same computer architecture.

The accelerations of the planets and asteroids differ by orders of magnitude, which means that normalization of the training data is essential to train the network. However, since HNNs have physics embedded into their architecture, we cannot normalize the inputs or outputs independently without breaking the physical constraints. For example, re-scaling the inputs between 0 and 1 implies that the relation between different inputs does not remain constant.  
The distributions of inputs and outputs have been included in \autoref{fig:distributioninputs} and \autoref{fig:distributionoutputs}, respectively.

\subsection{Architecture and training parameters}\label{subsec:hyperparam}

In order to make a fair comparison between the DNN and the HNN, the settings chosen will be common for both of them unless otherwise stated. Each of the two cases studied (SJS and SJSa) requires different neural network hyperparameters. For the SJS case, we adopt a Mean Squared Error (MSE) loss function as indicated in Equation ({\ref{eq:loss}}) for the HNN and Equation ({\ref{eq:lossDNN}}) for the DNN. For SJSa, the accelerations of the different bodies range multiple orders of magnitude, and therefore we implement a weighted MSE for the loss function, i.e, the error in the predicted acceleration of each body is weighted. The weights are applied to the losses defined in Equation({\ref{eq:loss}}) and Equation ({\ref{eq:lossDNN}}) as:

\begin{equation}\label{eq:lossweighted}
\mathcal{L}_{\text{NN}}(\theta) = 
W_1 \;\mathcal{L}_{a}(\theta)
+
W_2\;\mathcal{L}_{S}(\theta)
+
W_3\;\mathcal{L}_{J}(\theta),
\end{equation}
where $\mathcal{L}_a$, $\mathcal{L}_S$, and $\mathcal{L}_J$ represent the MSE loss for the accelerations of the asteroids, Saturn, and Jupiter, respectively. We empirically find that weights of $W_1=100$, $W_2 =10$, and $W_3=1$ produce the best results as these weight values relate to differences in orders of magnitude of the accelerations of the bodies. These weights are only necessary due to the impossibility of normalizing the inputs and outputs without breaking the physics constraints of the HNN. Although normalization is possible with the DNN, we have used the weighted loss function instead to get a fair comparison with the HNN.

For SJS, no hyperparameter optimization is carried out, but the architecture is chosen manually instead. Both the DNN and the HNN have three layers, 200 neurons in the first hidden layer and each hidden layer has 0.7 times the number of neurons of the previous one. The learning rate follows an exponential decay, with an initial learning rate of 0.01, a decay of 0.9, and $2\times 10^5$ steps. We use 150,000 samples with a proportion of 90/10 for training and validation datasets, and 10,000 samples for the test dataset.

\begin{figure}[h!]
	\centering
	\includegraphics[scale=0.36]{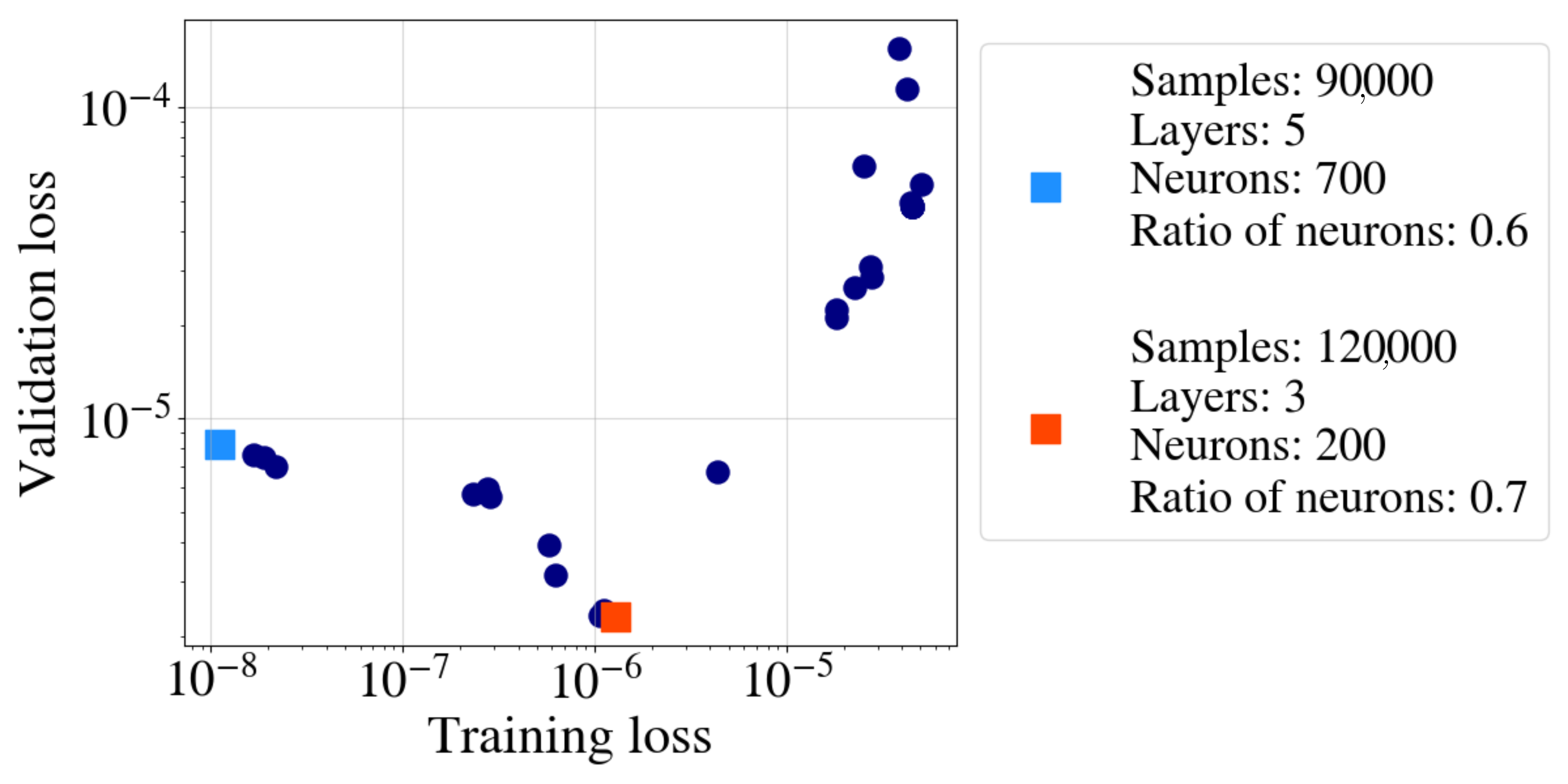}
	\caption[Results of the hyperparameter optimization.]{Results of the hyperparameter optimization for the Hamiltonian Neural Network. The training loss is plotted against the validation loss. Each point represents one trained network during the hyperparameter optimization. The points with the best training and validation loss are represented by blue and red squares, respectively, and their associated parameters are shown in the legend.}
	\label{fig:hyperparamOpt}
\end{figure}

For SJSa, the training of the HNN is not straightforward. To find a suitable combination of parameters, we perform a hyperparameter optimization where the variables are the number of training samples, number of layers, number of neurons per layer, and the learning rate parameters. We use a randomized grid search to explore different combinations of those parameters and train $\sim$30 networks for 200 epochs. The results are presented in \autoref{fig:hyperparamOpt}, where each simulation is plotted with the training loss along the $x$-axis and the validation loss along the $y$-axis. The figure indicates that regardless of the choice of parameters, the training and validation loss cannot be improved simultaneously to achieve the desired accuracy during testing. Among the best solutions, we choose the network architecture with three layers, 300 neurons per layer, and we use 250,000 samples for the training dataset. The test dataset is chosen to consist of 10,000 samples. The learning rate is chosen to follow an exponential decay schedule with an initial learning rate of $10^{-3}$, 800,000 steps, and a decay rate of 0.9. The same parameters are used for the DNN.

Some of the most commonly used activation functions fail to capture the characteristics of the problem. For example, the activation function has to take into account the large dynamic range of the values of the problem.  Therefore, we select the \texttt{SymmetricLog} activation function,
\begin{equation}\label{eq:simlog}
f(x) =\text{tanh}(x)\text{log}(x\text{tanh}(x) +1),
\end{equation}
which was specifically designed for this problem by Cai et al.\ in 2021 \cite{cai2021neural}, together with a Glorot weight initialization \cite{glorot2010understanding}.

\begin{figure}[h!]
	\centering
	\includegraphics[scale=0.45]{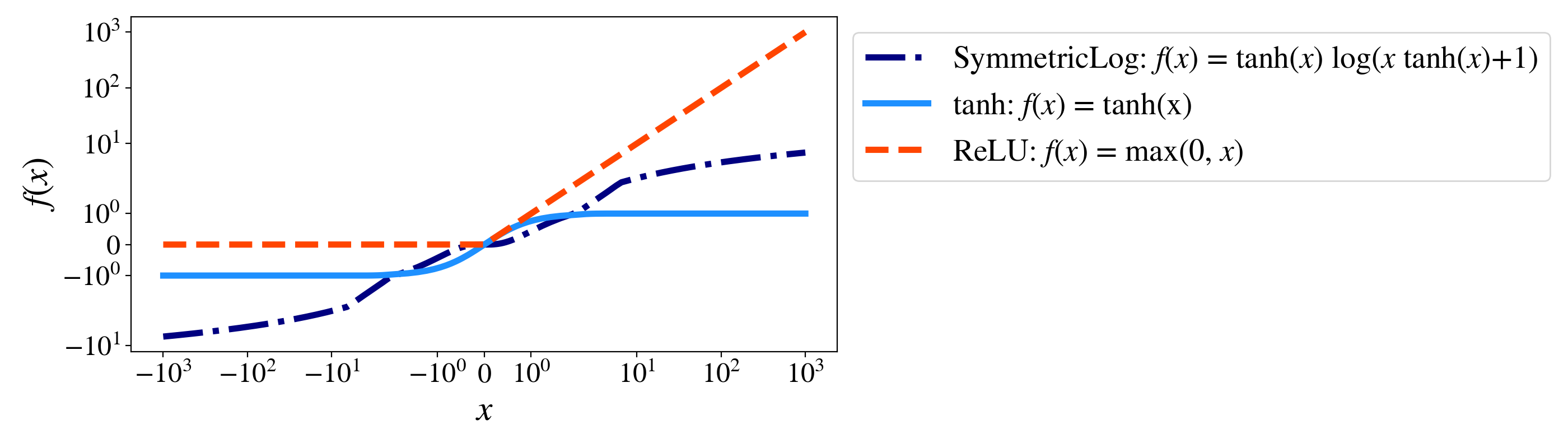}
	\caption[]{Comparison of activation functions. The SymmetricLog activation function was created for this problem by Cai et al.\ \cite{cai2021neural}.}
	\label{fig:tanh_log}
\end{figure}

This function behaves similarly to \texttt{tanh} close to zero, and like a logarithmic function for larger values. Moreover, it is symmetric for positive and negative values, as seen in \autoref{fig:tanh_log}.

\subsection{Training results}\label{subsec:trainresults}
Both the DNN and the HNN are trained using the Adam optimizer \cite{adam} for 2,000 epochs. For SJS, this takes $\sim$1.3 h and $\sim$2.3 h for the DNN and the HNN, respectively, on the same computer as we used for the creation of the dataset. For SJSa, the training time is $\sim$2.5 h and $\sim$5 h for the DNN and the HNN, respectively.

\begin{figure}[h!]
	\centering
	\includegraphics[scale=0.33]{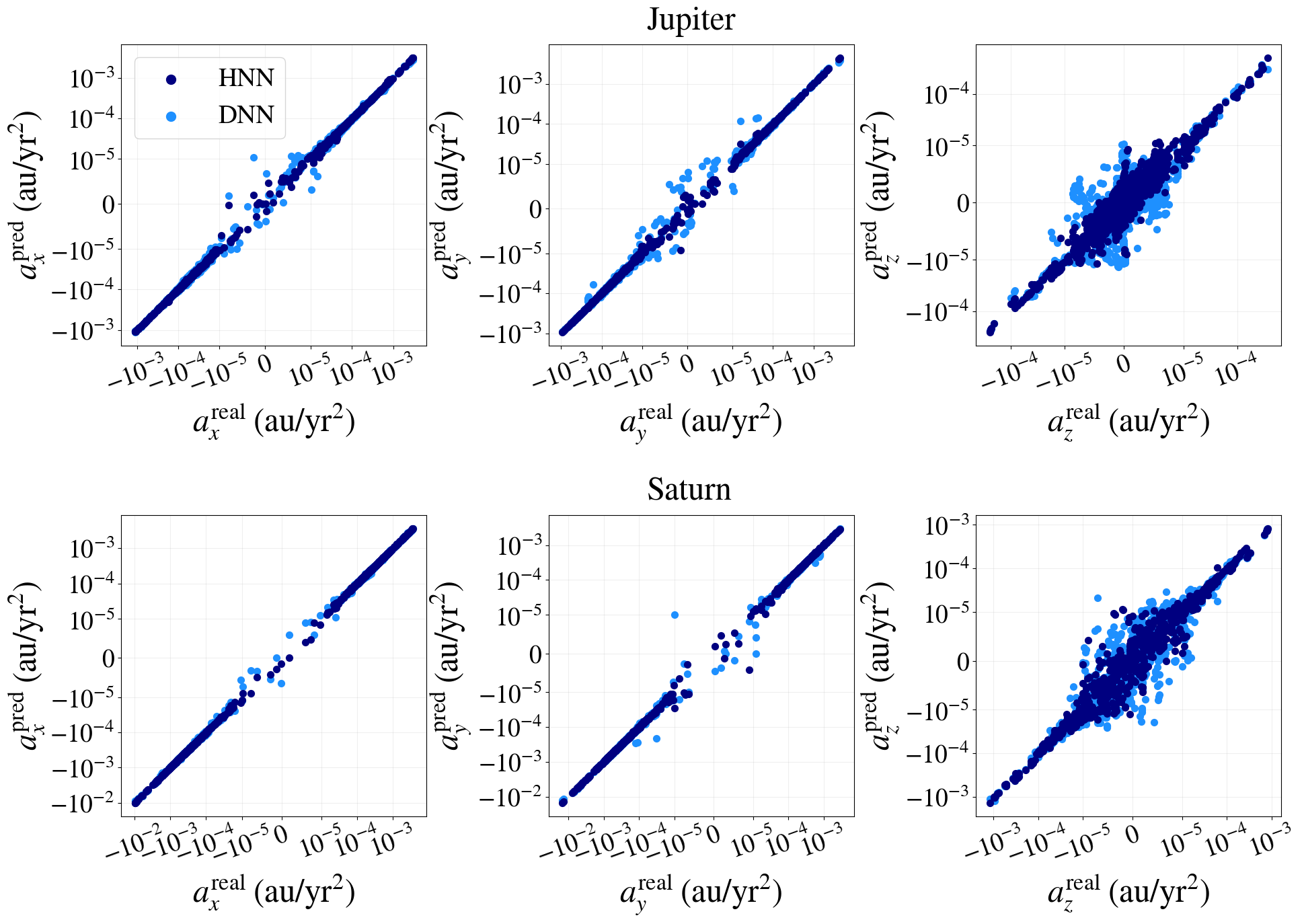}
	\caption[Error in prediction of the test dataset for the SJS]{Real against predicted values of the acceleration components for the case with the Sun, Jupiter, and Saturn. The real value is compared to the one predicted by the Deep Neural Network and the Hamiltonian Neural Network.}
	\label{fig:ErrorPred_SJS}
\end{figure}

Once the networks have been trained, we check their accuracy by applying them to the test dataset. For SJS, both the DNN and the HNN converge to a low loss value. \autoref{fig:ErrorPred_SJS} shows the prediction error for the accelerations obtained with each network. Both networks produce accurate results when the accelerations are large as their output is very close to the $45^{\circ}$ zero-error line. 
The relative error grows as the value of the acceleration decreases, 
since the absolute prediction error is in the order of $10^{-5}$. The errors of the DNN are larger than those of the HNN and overestimate the accelerations in the $y$-direction of Jupiter and in the $x$-direction of Saturn, and underestimate the $y$-acceleration of Saturn. This asymmetry leads to a drift in the energy error, as we will explain in \autoref{sec:results}. Due to the orbits being almost planar, the accelerations in the $z$-direction are smaller than for the $x$- and $y$-direction, which in \autoref{fig:ErrorPred_SJS} appears as a larger dispersion of small values.

\begin{figure}[h!]
	\hspace{-20pt}
	\includegraphics[scale=0.35]{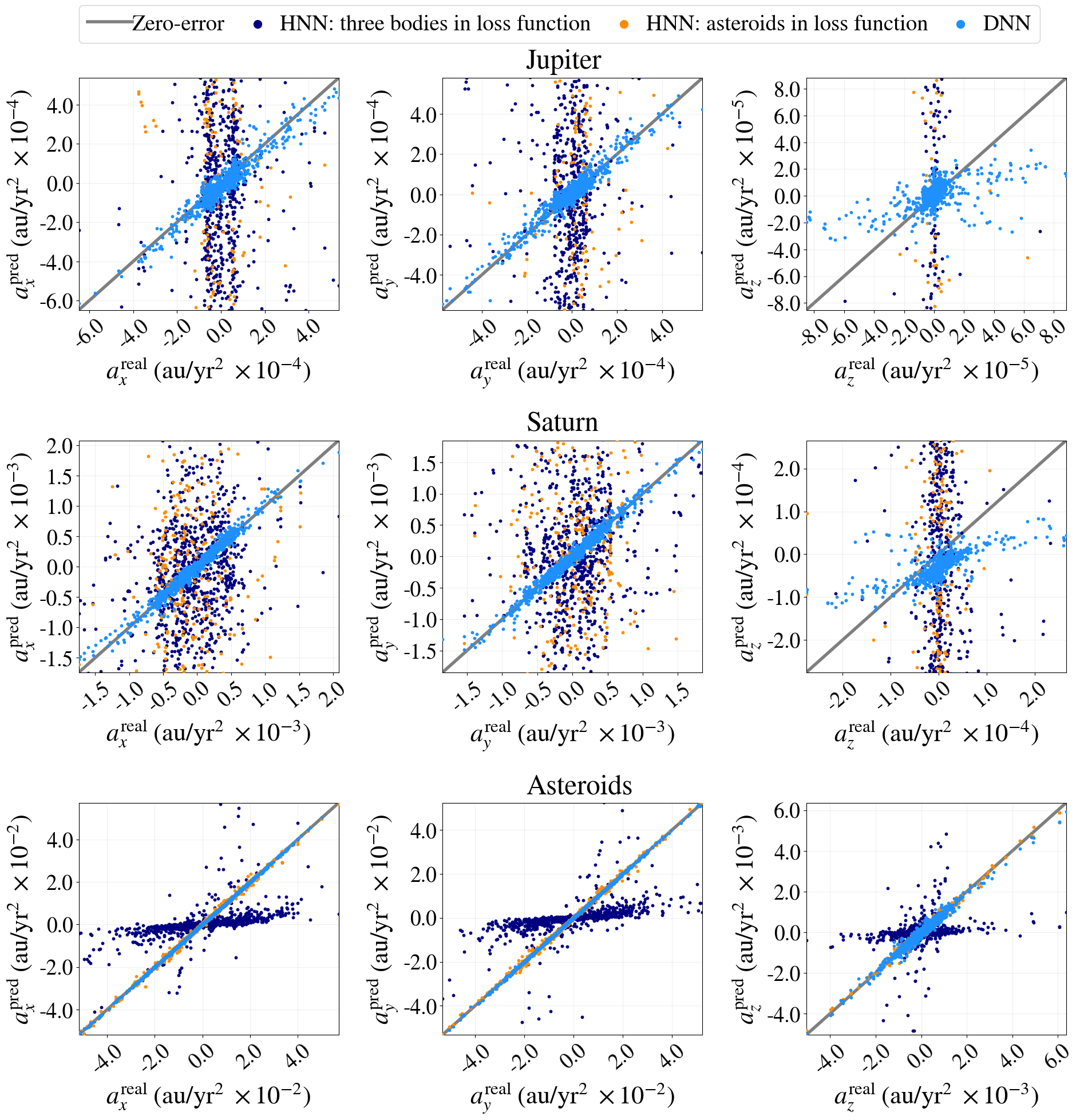}
	\caption[Error in prediction of the test dataset for the SJSa case with the Deep Neural Network and the Hamiltonian Neural Network.]{Real against predicted values of the acceleration components for the case with the Sun, Jupiter, Saturn, and asteroids. The real value is compared to those predicted by the Deep Neural Network, the Hamiltonian Neural Network for the three bodies, and the Hamiltonian Neural Network with only the asteroids in the loss function.}
	\label{fig:ErrorPred}
\end{figure}

The hyperparameter optimization in \autoref{subsec:hyperparam} shows that we fail to train the HNN for the SJSa case to a satisfactory loss value. Because of the large difference in masses between the asteroids and the planets ($\sim$7 orders of magnitude), when calculating the loss function, some of the gradients of the output with respect to the inputs are required to be extremely large, whereas others have to be small. This leads to the training process focusing on improving the predictions of the accelerations of the asteroids or the planets and, after a certain loss value is achieved, improving one of these implies making the others worse.
As a solution, since we successfully trained a network that predicts the accelerations of Jupiter and Saturn, we now train a network that solely predicts the accelerations of the asteroids. We therefore train another HNN where we only include the accelerations of the asteroids in the loss function, ignoring the predictions for Jupiter and Saturn. These results are presented in \autoref{fig:ErrorPred}. 

The DNN is trained with all the bodies in the loss function and can accurately predict the accelerations but, similarly to the predictions for SJS depicted in \autoref{fig:ErrorPred_SJS}, makes errors on the same side of the $45^\circ$ zero-error line. The HNN trained for the three bodies makes poor predictions for all the outputs, and the HNN trained only for the asteroids predicts the accelerations for the asteroids accurately but (as expected since they are not included in the loss function) fails to predict the accelerations for Jupiter and Saturn (\autoref{fig:ErrorPred}).

\subsection{Selection of networks}
For SJSa, the HNN fails to predict the accelerations of Jupiter, Saturn, and the asteroids simultaneously. However, if the network is trained with the loss only accounting for the prediction of the accelerations of the asteroids, it can predict these accurately as we discussed in \autoref{subsec:trainresults}. For SJSa, we will therefore calculate the accelerations using a combination of two networks: the predictions for Jupiter and Saturn with the network trained for SJS (\autoref{fig:ErrorPred_SJS}), and the prediction for the asteroids with the network that is only trained to predict the accelerations of the asteroids (orange markers in \autoref{fig:ErrorPred}). This combination of two networks is done with both the HNN and the DNN.

\subsection{Output of the HNN}
It is interesting to understand if the output of the HNN is the same as the actual interactive Hamiltonian of the system  (\autoref{eq:H_inter}). To test this hypothesis, we set up an experiment for SJS in which we compare the output of the HNN with the interactive energy of the system. 

\begin{figure}[h!]
	\centering
	\includegraphics[scale=0.35]{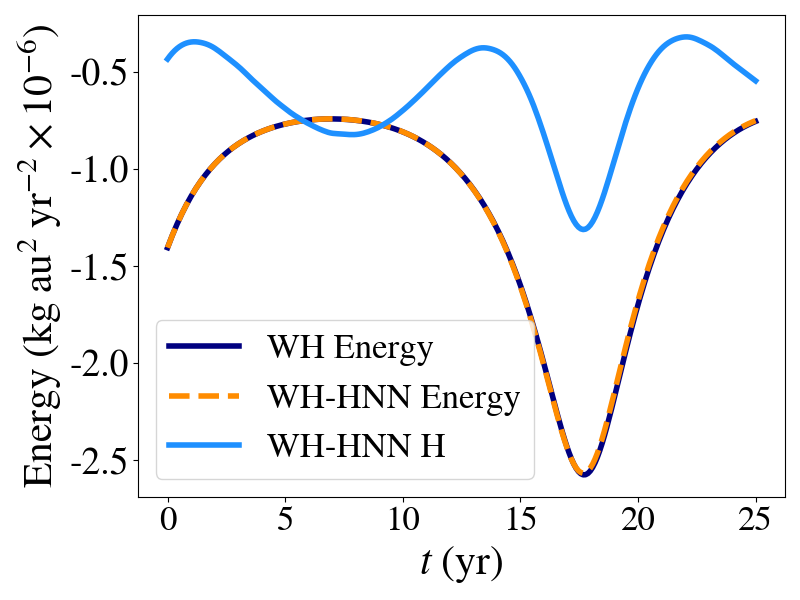}
	\caption[Comparison of the output of the HNN and the interactive Hamiltonian]{Comparison of the output of the Hamiltonian Neural Network with the interactive Hamiltonian of the Wisdom-Holman integrator. }
	\label{fig:HvsE}
\end{figure}

In \autoref{fig:HvsE}, we show that the predicted values of the interactive Hamiltonian with the HNN, i.e., $f_{\text{HNN}}(X)$ (WH-HNN H in \autoref{fig:HvsE}) are not the same as the interactive energy of the hybrid integrator $\mathcal{H}_{\text{inter}}^{\text{WH-HNN}}$ (WH-HNN Energy in \autoref{fig:HvsE}). The energy of the numerical solution $\mathcal{H}_{\text{inter}}^{\text{WH}}$ is also plotted as WH Energy for reference. 
The energy evolution of the hybrid integrator exactly coincides with the one of the numerical solution. The output of the HNN does not correspond to the energy value. Therefore, the output of the network does not have physical meaning. This can be explained by realizing that the accelerations obtained with the HNN depend on the relation between the output and the gradients. As a consequence, different combinations of these two variables may lead to similar values of the accelerations.

\section{Results of the hybrid Wisdom-Holman integrator} \label{sec:results}

In this section, we use the networks trained in \autoref{sec:ANNs} in a simulation to further study their performance.

\subsection{Integration parameters}
\label{subsec:Integrationparams}
We initialize the simulation with the state of the Sun, Jupiter, and Saturn from the Horizon System of the Jet Propulsion Laboratory \cite{jplephemeris}. We consider a variable number of asteroids initialized with a semi-major axis chosen randomly between 2.2 and 3.2 au, an eccentricity of 0.1, an inclination of $0^\circ$, and a random true anomaly. Then, we use the Wisdom-Holman integrator with a time step ($h$) of 0.1 yr until a final integration time which depends on the specific case (SJS or SJSa).

\subsection{Validation of the code}
\label{subsec:validation}
\begin{figure}[h!]
	\centering
	\includegraphics[scale=0.43]{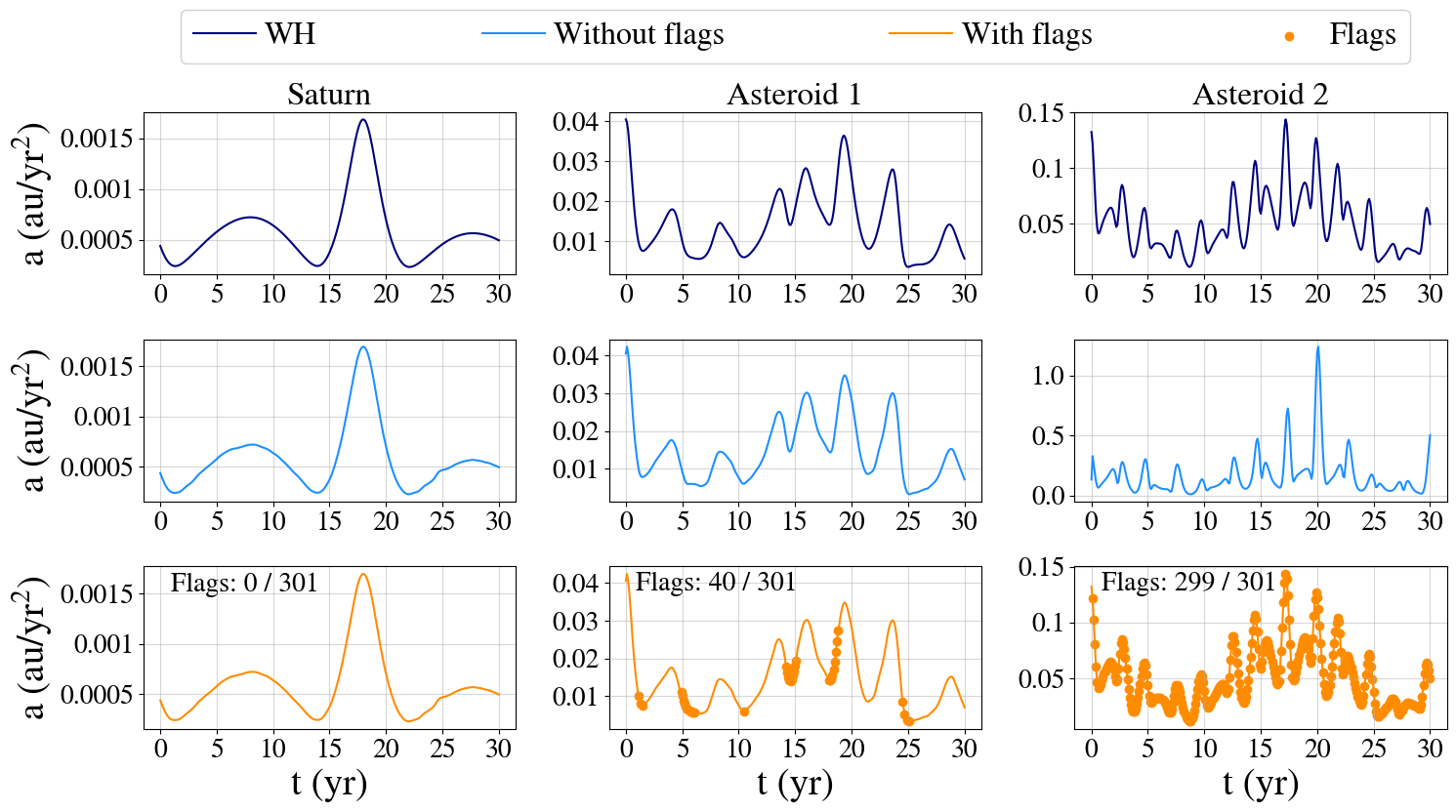}
	\caption[Comparison between the pure and hybrid integrators]{Comparison of the accelerations of Saturn, asteroid 1 with an orbit inside the range of the training dataset, and asteroid 2 with an orbit outside the range of training data, using different integration setups.  \textit{First row:} Wisdom-Holman integrator, \textit{second row:} Wisdom-Holman integrator with a Hamiltonian Neural Network, and \textit{third row:} hybrid Wisdom-Holman integrator with a Hamiltonian Neural Network and $R= 0.3$. In the third row, the dots represent the points in which the numerical integrator was used because the prediction of the neural network was not considered sufficiently accurate.}
	\label{fig:flagvsnoflag}
\end{figure}

Before discussing the results, we validate the hybrid implementation of the Wisdom-Holman integrator with the neural network. For this purpose, we compare two methods for SJSa: without replacing the HNN result by that of the numerical integrator if the requirement (Equation (\ref{eq:hybridcriteria})) is not achieved (without flags), and the method with flags as described in \autoref{sub:hybrid}. In \autoref{fig:flagvsnoflag} we show the accelerations of Saturn and two asteroids: asteroid 1 within the limits of the training dataset and asteroid 2 outside to study the extrapolation capabilities of the network. When the prediction of the network is accurate, as it is for Saturn, no flags are needed. However, when the network is not able to reproduce the numerical results, as in the case of asteroid 2, the hybrid integrator detects the poor predictions and replaces these with the results of the numerical calculation. By doing so, the hybrid HNN method becomes significantly more robust against prediction errors. In \ref{appendix:hybrid}, we discuss the number of flags as a function of the parameter $R$ from Equation (\ref{eq:hybridcriteria}).

We show in \autoref{fig:flagvsnoflag} that the hybrid integrator yields better solutions for the accelerations. However, verifying the predictions of the networks at each time step entails a cost in terms of computing time. 

\begin{figure}[h!]
	\centering
	\includegraphics[scale=0.41]{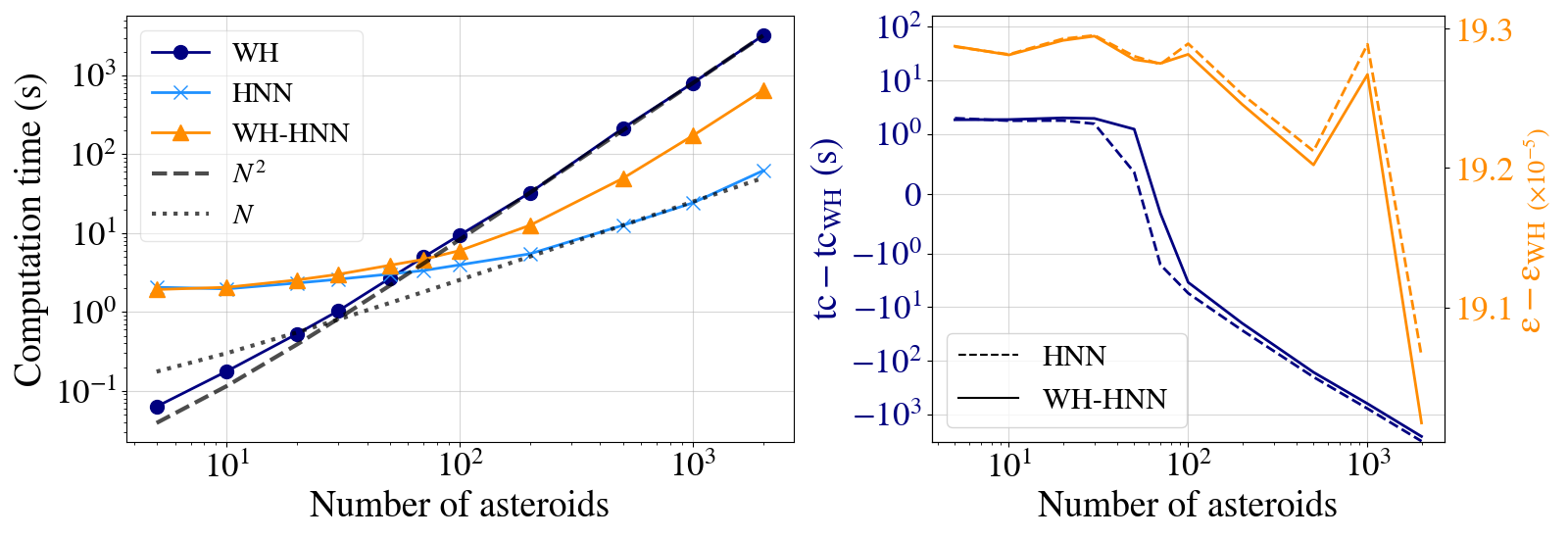}\\
	$ $\hspace{80pt}(a) \hspace{190pt} (b) \hspace{60pt}
	\caption[]{Computing time (a), and difference in computing time and energy error with respect to the numerical solution (b) for the integration to 20 years as a function of the number of asteroids. Three cases are shown: numerical integrator (WH), numerical integrator with Hamiltonian Neural Network (HNN), and hybrid numerical integrator with Hamiltonian Neural Network (WH-HNN). The $N$ and $N^2$ lines are displayed as a reference for linear and quadratic scaling, respectively.}
	\label{fig:timeEnergyAsteroids}
\end{figure}

The numerical integration scales with $N^2$ whereas the neural network result scales with $N$. For a small number of asteroids, the additional computing time needed to include the neural networks into the integrator makes the method with neural networks more expensive than the numerical computation. We therefore study what the minimum number of asteroids is to make the use of neural networks computationally less expensive than the numerical computation. In \autoref{fig:timeEnergyAsteroids}, three cases are displayed: Wisdom-Holman integrator, WH with HNN without flags, i.e. HNN, and hybrid WH with HNN, i.e, WH-HNN. For a number of asteroids $\leq$70, the use of the HNNs is not preferred above WH as it takes longer to run. However, as the number of asteroids increases, 
using either HNNs or the hybrid method with HNNs within the integrator results in faster computations, halving the computing time for 2,000 asteroids. Using the hybrid method with the HNN only slightly increases the computing time with respect to the pure HNN case since the prediction for each asteroid is evaluated and replaced individually if necessary. In \autoref{fig:timeEnergyAsteroids}b, we see that the hybrid integrator reduces the energy error without significantly increasing the computing time. Since the energy error is dominated by the planets, 
a small improvement in the energy error implies a significant improvement in the predictions of the accelerations of the asteroids. 

The computing times shown in \autoref{fig:timeEnergyAsteroids} refer to the times for the calculation of the accelerations, i.e., the training times for the neural networks are not included. Once the networks are trained, they can be used in multiple experiments. For example, if the objective is to run 100 experiments, a training time of 2 h is negligible compared to the total computing time.

\subsection{Trajectory integration}\label{subsec:results}
\begin{figure}[h!]
	\centering
	\hspace{-25pt}
	\includegraphics[scale=0.34]{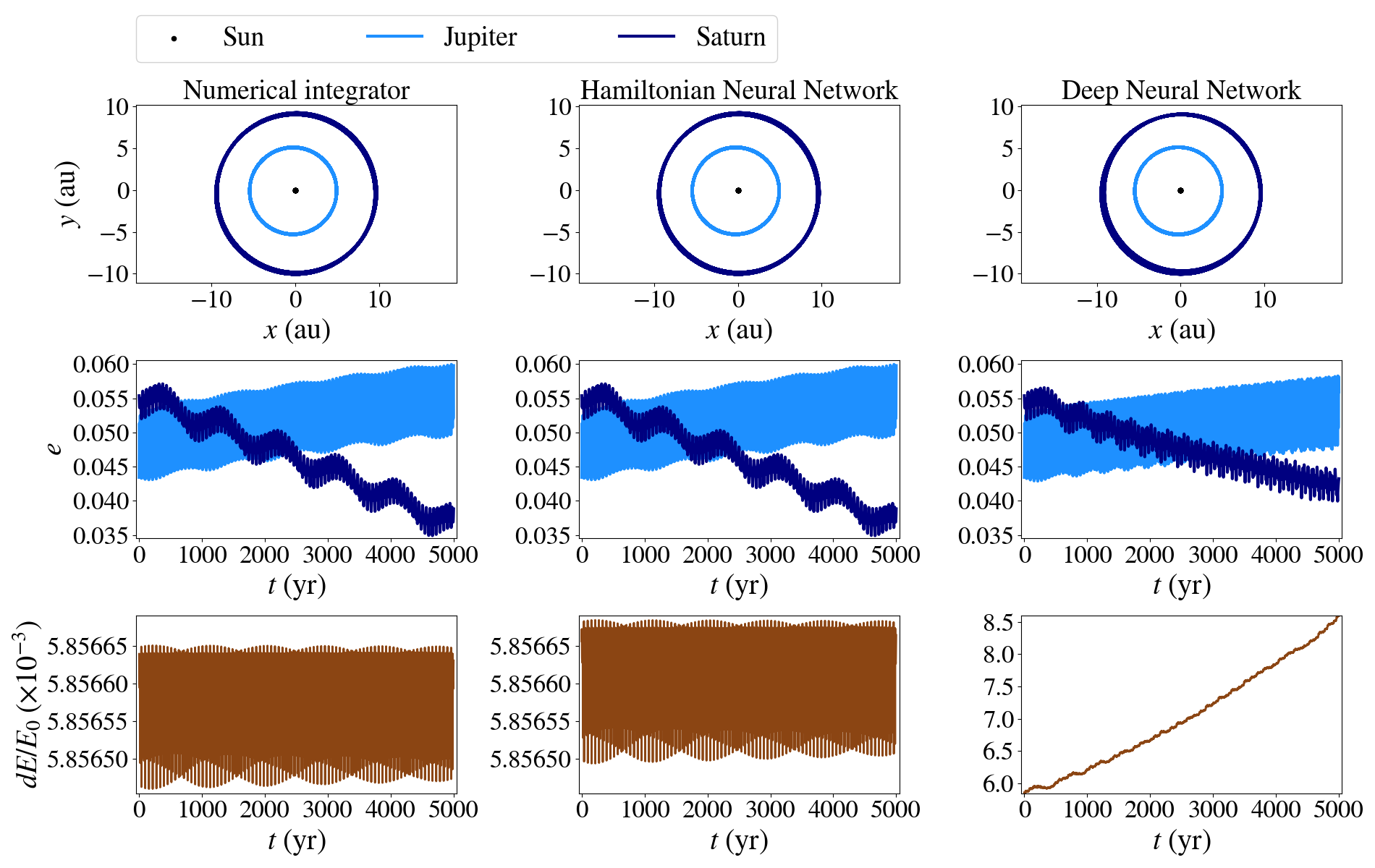}
	\caption[Simulation results for the SJS case]{Simulation results for the Sun, Jupiter, and Saturn (SJS). The results are generated using the Wisdom-Holman integrator (\textit{left}), hybrid Wisdom-Holman with the Hamiltonian Neural Network (\textit{middle}), and hybrid Wisdom-Holman with the Deep Neural Network (\textit{right}). The trajectories in the $x$-$y$ plane are shown in the first row, the eccentricities in the second row, and the relative energy error in the third row. 
	}
	\label{fig:resultSJS}
\end{figure}

\begin{figure}[h!]
	\centering
	\hspace{-20pt}
	\includegraphics[scale=0.33]{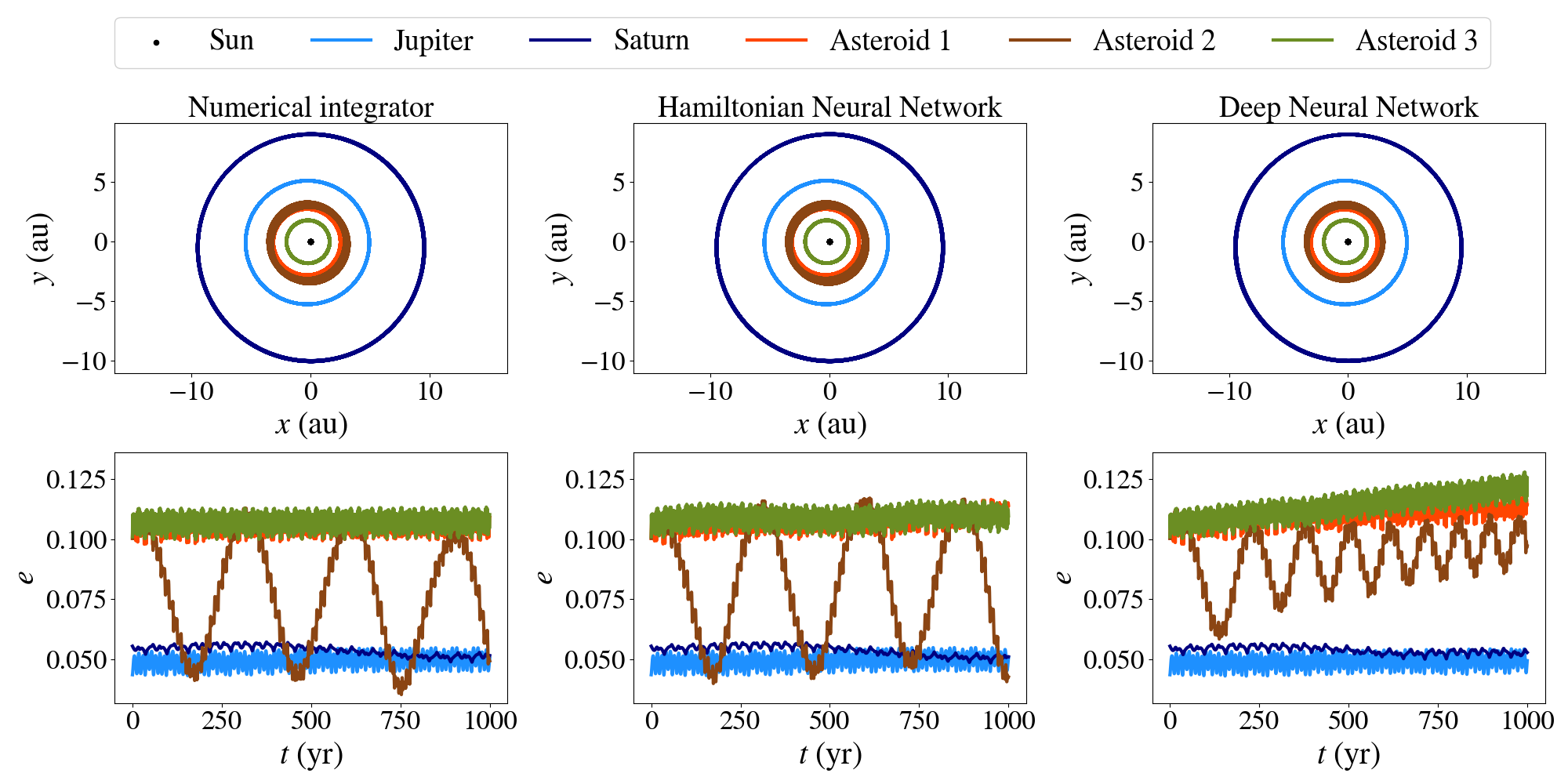}
	\caption[Simulation results for the SJSa case]{Simulation results for the case with the Sun, Jupiter, Saturn, and asteroids. The results are generated with the Wisdom-Holman integrator (\textit{left}), hybrid Wisdom-Holman with the Hamiltonian Neural Network (\textit{middle}), and hybrid Wisdom-Holman with the Deep Neural Network (\textit{right}). The trajectory in the $x$-$y$ plane is shown in the first row and the eccentricity in the second row. Asteroids 1 and 2 are within the limits of the training dataset and asteroid 3 has a semi-major axis below the lowest limit of the training dataset to study the extrapolation capabilities of the networks. }
	\label{fig:resultSJSa}
\end{figure}

Once the neural networks have been trained, we integrate SJS for 5,000 years (\autoref{fig:resultSJS}) and SJSa for 1,000 years (\autoref{fig:resultSJSa}). To study the extrapolation capabilities of the network, we add two asteroids to SJSa, of which the initial conditions are within the range of training parameters (asteroids 1 and 2) and one asteroid with a semi-major axis outside the range  (asteroid 3). 

In \autoref{fig:resultSJS}, we compare the trajectory, change in eccentricity, and energy error of the hybrid integrator with the HNN and the DNN with respect to the numerical integration. The integrator with the HNN (WH-HNN) reproduces the change in eccentricity of the integrator better than the one with the DNN (WH-DNN). The evolution of the eccentricity is an important indicator of how well the orbit is reproduced using the neural networks. Another indicator is the energy error (third row). Although the WH-HNN leads to a larger energy error than the WH integrator, it shows symplectic behavior. In contrast, the use of the WH-DNN leads to a systematic drift in the energy error. This causes a gradual divergence from the numerical solution. We illustrate this with \autoref{fig:ErrorPred_SJS}, where the DNN produces prediction errors that are asymmetrically distributed around the zero-error line. We conclude that for the SJS case the hybrid integrators with the HNN and the DNN can reproduce the numerical results for short time scales, although the use of the latter results in a systematic deviation from the reference solution.

For SJSa, the results in \autoref{fig:resultSJSa} show that the trajectories of asteroids 1 and 2 can be predicted with both the WH-HNN and the WH-DNN for short integration times. For longer integration times, the DNN is not able to reproduce the trajectories of the asteroids accurately; there is a systematic drift in the evolution of the eccentricity. 
Regarding the extrapolation capabilities of the networks, neither the HNN nor the DNN can predict the trajectory of asteroid 3. However, the hybrid integration allows the accelerations to be adjusted to the numerical values, leading to more accurate trajectories. Regarding the energy error, the behavior observed is the same as in \autoref{fig:resultSJS} since the energy magnitudes of Jupiter and Saturn dominate over the energy magnitudes of the asteroids. We conclude that for short time scales both networks incur in a small error with respect to the numerical integration results, but the HNN achieves a more accurate reproduction of the trajectory of the asteroids over a longer time scale.

\section{Conclusion}\label{sec:conclusions}

In this paper, we studied the use of Artificial Neural Networks for the prediction of accelerations in a planetary system with a star orbited by two planets and a number of asteroids. We compared the results produced by a Deep Neural Network and a Hamiltonian Neural Network. The latter includes physical knowledge about the conservation of energy. 

In contrast to previous studies that use neural networks for the gravitational \textit{N}-body problem, we focused on an actual astrophysics problem. By using a case-specific integrator and modifying the number of bodies and their masses and positions to represent a realistic scenario, we encountered challenges that are not found when using this problem as a test case.

We created a method that circumvents some of the major challenges of using neural networks for the $N$-body problem. First of all, by using a hybrid integrator that evaluates the prediction and chooses between the numerical or the neural network solution, we addressed the problem of accumulation of errors over large timescales. Secondly, our setup allows for a variable number of bodies in the system without the need to retrain the network. With the simplest setups found in literature, an increase in the number of bodies in the system implies that the network needs to be retrained. Finally, we use custom activation functions and weights in the loss function to adapt to the characteristics of the problem.

Although based on the optimistic results from the literature \cite{Raissi2019, Lu2021, greydanus2019hamiltonian} we expected the HNN to outperform the DNN, in the case with the asteroids, the HNN could not be trained to predict simultaneously the accelerations of the planets and the asteroids. Because of the presence of physics constraints in HNNs, normalization is not possible. This becomes an obstacle for training due to the differences in masses of the bodies. We therefore trained two individual networks for the accelerations of the planets and the asteroids. Although using HNNs has its advantages for the simplified case with Jupiter and Saturn, we demonstrated their limitations for other configurations. 

HNNs turn out to be more time-consuming and harder to train, in contrast to the DNN. We had to develop a dedicated activation function specifically for this problem and the hyperparameter optimization performed was time-consuming as well. 

With more than 70 asteroids, the integration with the neural networks becomes faster than the direct numerical integration, and for 2,000 asteroids the use of neural networks leads to a halving of the computing time. Since the goal is to create a method that can be used multiple times, the performance comparison does not include the time used for training.

We developed a hybrid integrator to alleviate the problems induced by the introduction of neural networks in the integration process.
By verifying the prediction made by the ANN at each time step and replacing this prediction by the numerical integrator if necessary, the integrator becomes more reliable and robust to prediction errors without significantly increasing the computing time. Therefore, for a sufficiently large number of asteroids ($\sim70$), we find that the hybrid approach with the HNN proposed here outperforms the direct integration without losing the underlying physics of the system, as opposed to the hybrid integrator with the DNN. Although our study shows that it is beneficial to use physics-aware architectures that conserve the symplectic structure of the integrator, our hybrid method is independent of the network topology chosen. We focused on the simplest cases of neural networks to allow for a better understanding of the underlying challenges of the problem, but further studies should focus on the use of more complex network topologies. 

In short, we showed that neural networks can be used to speed up the integration process for problems with a large number of asteroids. However, for long integration times, the prediction errors may accumulate causing the results to diverge with respect to the solution obtained by direct numerical integration. Moreover, if no hybrid integration method that verifies the prediction of the network is used, these prediction errors may lead to unphysical solutions on a short time scale. The use of HNNs is justified for cases in which normalization is not needed to train the network, which in this study means when the masses of the different bodies are of the same order of magnitude. When the HNNs can be trained, they show symplectic behavior, with the energy error oscillating around the initial value. In contrast, DNNs are easy to train and lead to satisfactory solutions, but are not able to extrapolate to conditions that are not part of the training data and are unsuited for finding solutions that conserve energy.

\section{Acknowledgments}
This publication is funded by the Dutch Research Council (NWO) with project number OCENW.GROOT.2019.044 of the research programme NWO XL. It is part of the project “Unravelling Neural Networks with Structure-Preserving Computing”.
In addition, part of this publication is funded by the Nederlandse Onderzoekschool Voor Astronomie (NOVA).

\newpage

\bibliographystyle{ieeetr}
\bibliography{references}

\appendix
	\section{Dataset parameters}\label{appendix:datasetparam}
	
	The parameters for the simulation and initial conditions are shown in \autoref{table:dataset_inputs}. The orbital elements which have not been included in the table, i.e., the right ascension of the ascending node and the argument of the periapsis, have been set to zero. 
	\begin{table}[h!]
		\begin{center}
			\caption[]{Summary of dataset parameters. sma is the semi-major axis. J represents Jupiter, S Saturn and a the asteroids}
			\small
			\begin{minipage}{0.5\textwidth}
				\begin{tabular}{| l |c  | }
					\hline
					Parameter
					&SJS \& SJSa\\
					\hline
					Experiments (train) & 500 \\
					Experiments (test) & 25 \\
					Time step & 5.0 $\times 10^{-3}$ yr\\
					Final time & 30 yr\\
					Mass J ($m_J$) & 9.543e-4 $M_{\text{Sun}}$\\
					Mass S ($m_S$) &2.857e-4 $M_{\text{Sun}}$\\
					Mass a ($m_a$)&  [$1\times 10^{19}$, $1\times 10^{20}$] kg\\
					sma J ($a_J$) & [4, 8] au\\
					sma S ($a_S$)& [8.5, 10] au \\
					sma a ($a_a$)& [2.2, 3.2] au\\
					\hline
				\end{tabular}
				
			\end{minipage}
			\begin{minipage}{0.45\textwidth}
				\begin{tabular}{| l |c  | }
					\hline
					Parameter
					&SJS \& SJSa\\
					\hline
					Eccentricity J ($e_J$)& [0, 0.1]\\
					Eccentricity S($e_S$)& [0, 0.1]\\
					Eccentricity a ($e_a$)& [0, 0.1]\\
					Inclination J ($i_J$) & [0, $6^\circ$] \\
					Inclination S ($i_S$) & [0, $6^\circ$] \\
					Inclination a ($i_a$) &[0,  $6^\circ$]\\
					True anomaly J ($f_J$)  & [0, $360^\circ$]\\
					True anomaly S ($f_S$)& [0, $360^\circ$]\\
					True anomaly a ($f_a$)& [0, $360^\circ$]\\		
					%		cell4 & cell5 & cell6 \\  
					%		cell7 & cell8 & cell9   
					&\\ 
					\hline
				\end{tabular}
				
				\label{table:dataset_inputs}
			\end{minipage}
		\end{center}
	\end{table}

\begin{figure}[h!]
		\centering
		\includegraphics[scale=0.55]{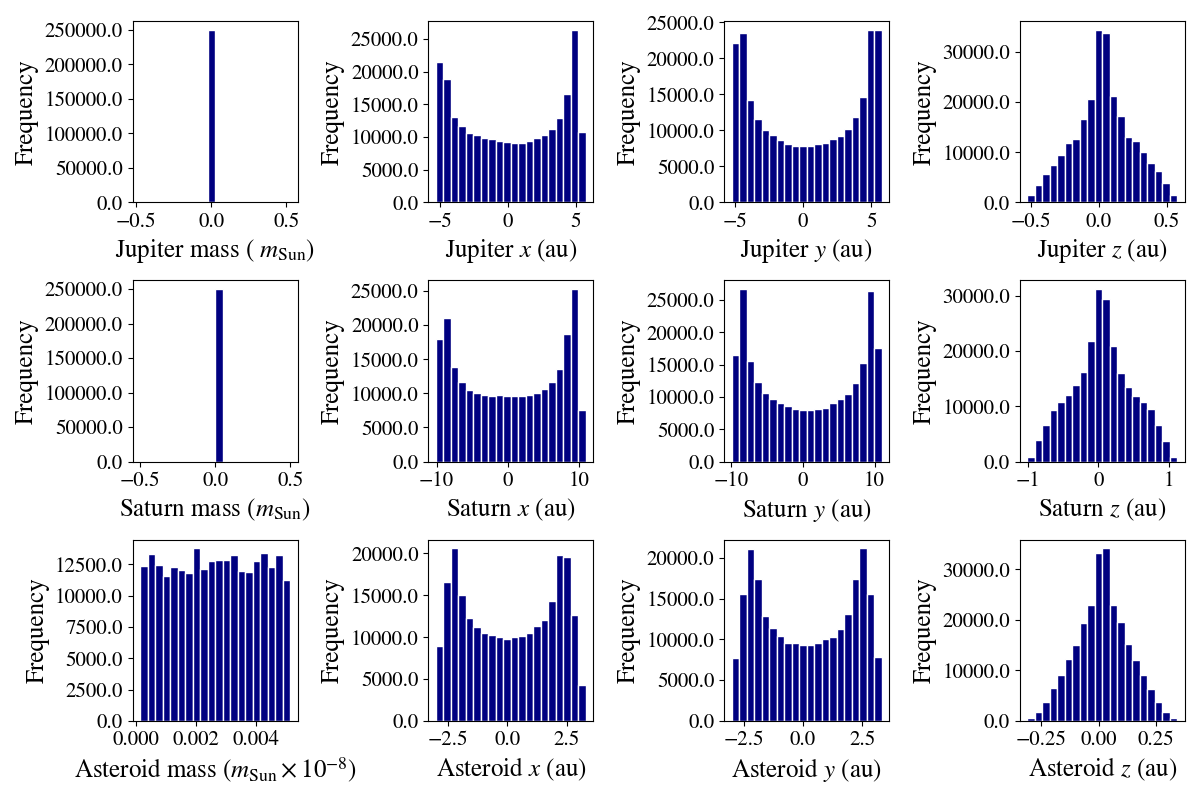}
		\caption[Distribution of inputs for the training dataset.]{Distribution of inputs for the training dataset. The inputs include the position vectors and masses of the two planets and the asteroids }
		\label{fig:distributioninputs}
	\end{figure}

\begin{figure}[h!]
		\centering
		\includegraphics[scale=0.55]{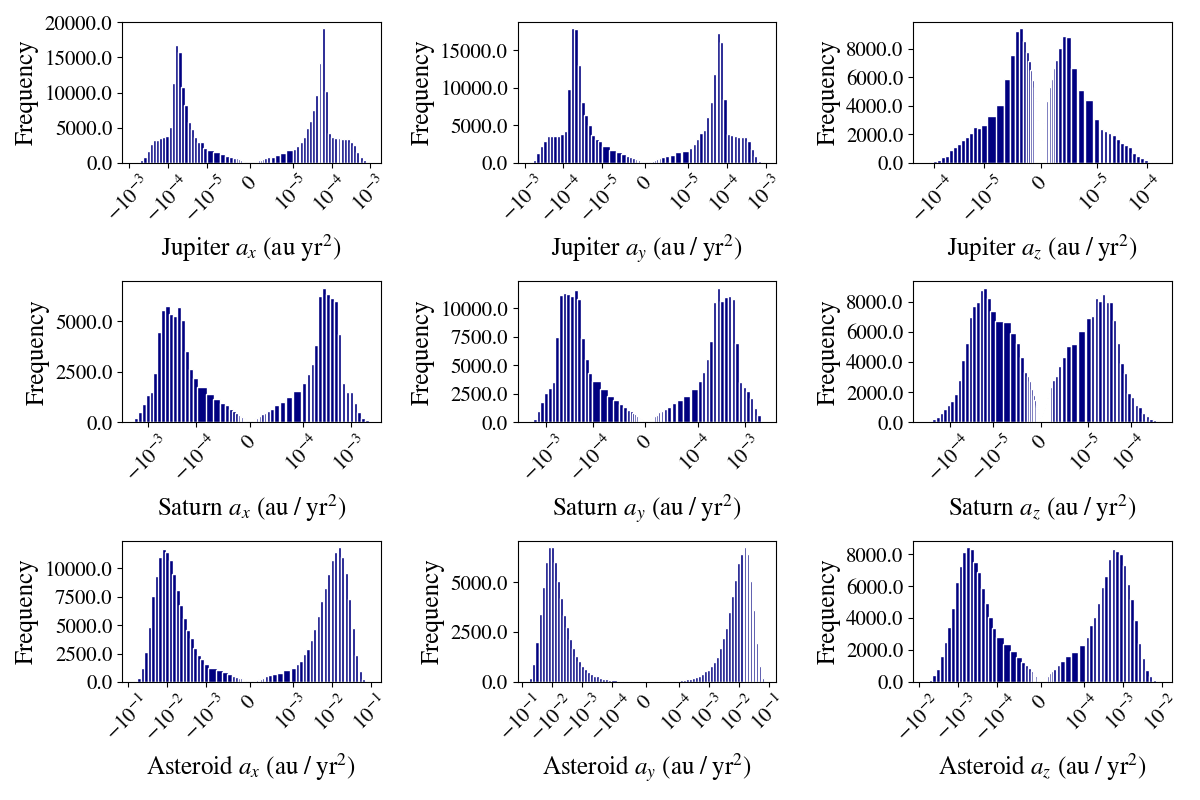}
		\caption[Distribution of outputs for the training dataset.]{Distribution of outputs for the training dataset. The outputs include the acceleration vectors of the two planets and the asteroids.}
		\label{fig:distributionoutputs}
	\end{figure}

	The distribution of inputs and outputs of the training dataset for the case with the Sun, Jupiter, Saturn, and the asteroids is shown in
	\autoref{fig:distributioninputs} and \autoref{fig:distributionoutputs}, respectively.
	
	
	\newpage
	\section{Hybrid method}\label{appendix:hybrid}
\begin{figure}[h!]
		\centering
		\includegraphics[scale=0.42]{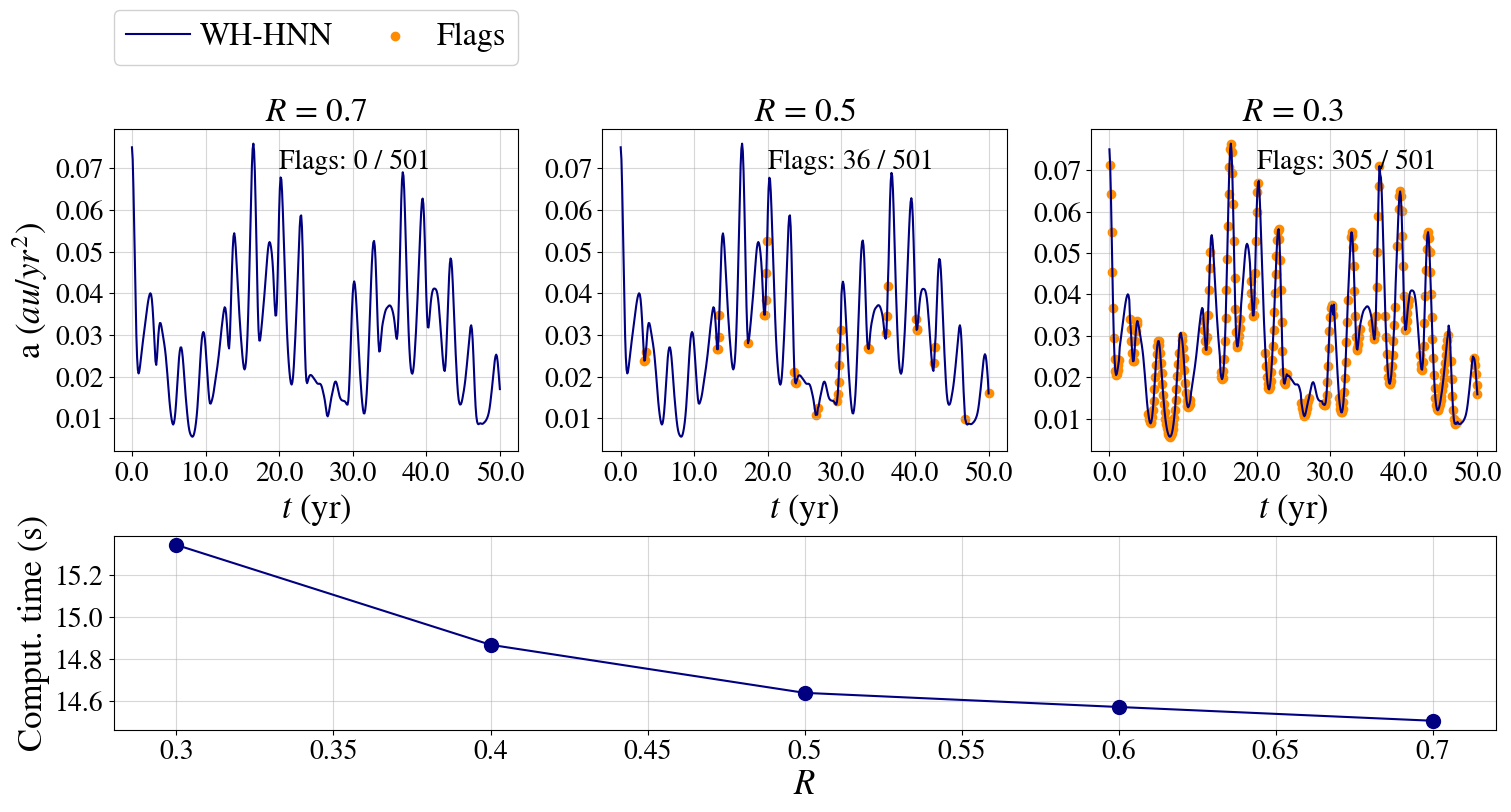}
		\caption[Comparison of R vs flags]{Comparison of the number of flags in the integration of Asteroid 1 with different values of $R$ (Equation (\ref{eq:hybridcriteria})). \textit{First row:} accelerations with flags for three different values of $R$. \textit{Second row:} Computing time as a function of $R$.}
		\label{fig:flagvsnoflagR}
	\end{figure}

	In Equation (\ref{eq:hybridcriteria}), we showed the criterion for rejecting the prediction of the neural network. Depending on the value of $R$ chosen, the balance between energy error and computing time changes. We show in \autoref{fig:flagvsnoflagR} the number of flags for three values of $R$ for the integration of asteroid 1 with the HNN. As $R$ increases, the method becomes less strict and the accelerations predicted are further from those calculated numerically. If $R$ is 0.1, all HNN values are rejected, and the whole simulation is done ignoring the predictions of the HNN.
	
	The computing time decreases as the number of flags needed is reduced, i.e, for larger values of $R$. 

\end{document}